# Development of a Deep Learning System for Intra-Operative Identification of Cancer Metastases

Thomas Schnelldorfer[1,2,3 ✉], Janil Castro[3], Atoussa Goldar-Najafi[4], Liping Liu[5]

[1]Division of Surgical Oncology, Tufts Medical Center, Boston, MA, USA.
[2]Department of Translational Research, Lahey Hospital and Medical Center, Burlington, MA, USA.
[3]Surgical Imaging Lab, Tufts Medical Center, Boston, MA, USA.
[4]Department of Pathology, Lahey Hospital and Medical Center, Burlington, MA, USA.
[5]Department of Computer Sciences, Tufts University, Medford, MA, USA.

✉email: thomas.schnelldorfer@tufts.edu

## Abstract

For several cancer patients, operative resection with curative intent can end up in early recurrence of the cancer. Current limitations in peri-operative cancer staging and especially intra-operative misidentification of visible metastases is likely the main reason leading to unnecessary operative interventions in the affected individuals. Here, we evaluate whether an artificial intelligence (AI) system can improve recognition of peritoneal surface metastases on routine staging laparoscopy images from patients with gastrointestinal malignancies.  In a simulated setting evaluating biopsied peritoneal lesions, a prototype deep learning surgical guidance system outperformed oncologic surgeons in identifying peritoneal surface metastases.  In this environment the developed AI model would have improved the identification of metastases by 5% while reducing the number of unnecessary biopsies by 28% compared to current standard practice.  Evaluating non-biopsied peritoneal lesions, the findings support the possibility that the AI system could identify peritoneal surface metastases that were falsely deemed benign in clinical practice.  Our findings demonstrate the technical feasibility of an AI system for intra-operative identification of peritoneal surface metastases, but require future assessment in a multi-institutional clinical setting.

## Introduction

Treatment selection for most cancers fundamentally relies on the absence or presence of grossly detectable distant metastases (i.e. staging) [1].  The accuracy of conventional staging through cross-sectional radiographic imaging is for several gastrointestinal malignancies considered limited, mostly due to the inability to identify smaller, yet significant distant metastases [2-5].  This particularly affects patients who are considered for operative resection of a presumed localized cancer.  As a result of inadequate staging, between 12% and 33% of patients with cancers of the pancreas, stomach, esophageal, small intestine, and bile duct/gallbladder who underwent operative resections with curative intent will end up dying of cancer progression within 1 year after the operation [6-13].  This substantial subgroup of patients, therefore, underwent an operation that can be morbid, with a temporary decrease in quality of life, increase in health care cost, and burden on patient families, but in retrospect without significant clinical or oncologic benefit and implicit delay of the more appropriate treatment of chemotherapy [14].  Hence, there is a great need for improved staging to allow for better treatment allocation of various cancer patients.

The above clinical observations indicate that distant metastases of not insignificant size must have been present in a great number of patients at the time of their cancer operation, but were either not visible to current imaging techniques or misclassified [15,16].  In particular, missed peritoneal surface metastases are likely a major cause for the observed high rate of early postoperative cancer deaths.  The peritoneum is one of the most common sites for metastases from gastrointestinal malignancies, but it is also one of the most difficult sites to detect smaller metastases on radiographic imaging [17].  Operative staging, and in particular staging laparoscopy, seems a logical solution.  Especially since the voxel for staging laparoscopy is 10-times smaller than the voxel for the most sensitive radiographic methods (i.e. staging laparoscopy can pick up much smaller changes than cross-sectional radiographic imaging) [18,19]. However, the problem encountered with operative staging is that benign peritoneal surface abnormalities (i.e. lesions) are prevalent and can mimic the appearance of metastases [16]. Biopsy with frozen section histopathology of every lesion in every patient, due to time constraints, is in clinical practice not feasible. Hence, surgeons are forced to make a subjective decision about when to pursue a biopsy of what is believed to be a suspicious-appearing lesion and when to consciously ignore what is presumed to be a benign-appearing lesion. Recent data from our working group showed that such subjective determination is very inaccurate [16]. Expert surgeons on average misidentified 36±19% of gross visible metastases, and simple macroscopic optical features, such as shape and color, are too similar between benign lesions and metastases to make a reliable classification [16].

Out of the above reasons, staging laparoscopy combined with an automated computer vision approach to classify visible changes of the peritoneum encountered during cancer operations could provide a solution. The aim of this study was to develop and test a deep learning prototype that can be used to identify peritoneal surface lesions and predict its pathology from routine staging laparoscopy images as a basis for future development of a computer-assisted staging laparoscopy (CASL) system. An illustration of the pipeline for development of the system and testing of its performance is outlined in Figure 1.

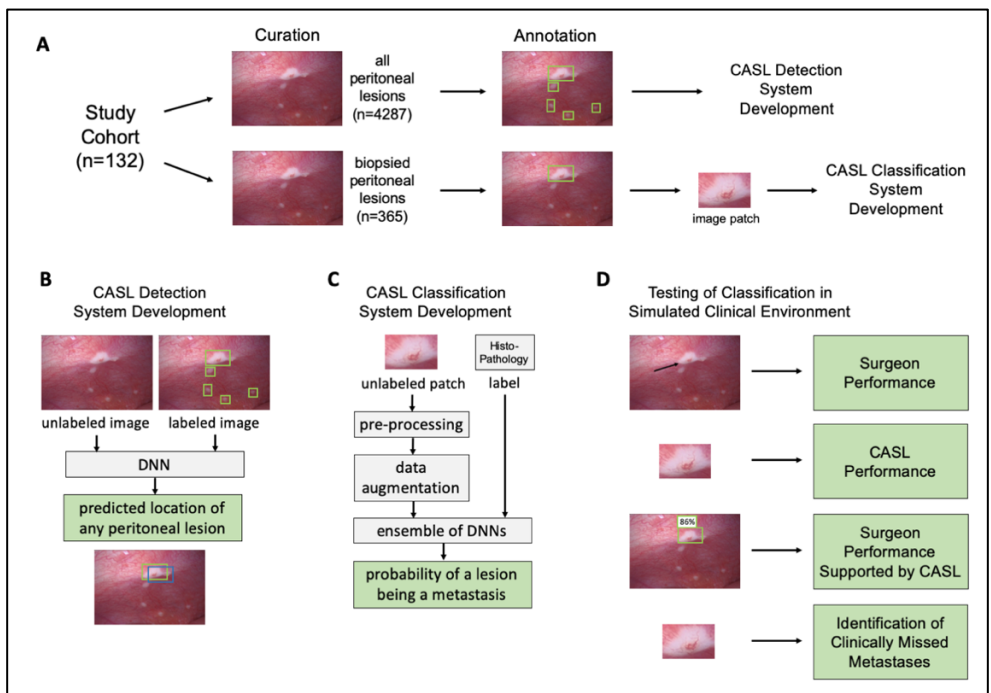

Figure 1: Model development and testing pipeline. A) Of 132 patients with gastrointestinal cancers who underwent staging laparoscopies with biopsies of peritoneal lesions, representative images of peritoneal surface lesions were curated and subsequently annotated according to location and, if biopsied, histopathology results representing the data used to develop CASL. B) A deep neural network (DNN) was developed to predict the location of any lesion on laparoscopy images. C) Random ensembles of DNNs were developed to predict the probability of malignancy in image patches of biopsied peritoneal surface lesions. D) Oncologic surgeons' predicted the probability of malignancy of peritoneal surface lesions depicted in the study's laparoscopy images with and without the support of CASL's classification system allowing for performance comparison of surgeon vs. CASL vs. surgeon with CASL in a simulated clinical environment. In order to assess the potential for CASL to identify clinically missed peritoneal surface metastases, the performance of CASL's classification system on images from patients that were not deemed to have distant metastases was tested.

## Results

**Study Demographic**
A total of 132 patients were entered into this study, including 89 men and 43 women with a mean age of 70 +/- 10 years. The study cohort included patients with adenocarcinoma of the stomach (n=57), pancreas (n=49), bile duct or gallbladder (n=12), small intestine (n=10), and other gastrointestinal sites (n=4). A total of 4287 peritoneal surface lesions were visualized (32.5 +/- 35.0 lesions per patient), including 365 lesions that were biopsied (2.8

+/- 1.6 biopsied lesions per patient). Biopsied lesions involved the parietal peritoneum (n=322), liver surface (n=26), and other visceral peritoneum sites (n=17). Thirty-three percent of biopsies were deemed malignant on final pathology. For those with positive peritoneal biopsy, the mean peritoneal cancer index was 7 +/- 5. Of all videos, 87% were depicted at a resolution of 1920x1080 pixels or above and all were depicted at a frame rate of 30 fps with a bit depth of 8 bits or above.

**CASL Detection System**

Early experiments have demonstrated that classification of peritoneal lesions on the entire image is very challenging given the heterogeneity of operative images. We, therefore, elected an approach to develop a detection system that would outline the lesion for subsequent classification by a separate system.

A detection model was trained and tested on 1092 images depicting all 4287 lesions seen from 132 patients (i.e. 'all lesion dataset'). The location of each lesion was marked by a tight bounding box around each lesion on labeled images. The model was trained to autonomously find these lesions on unlabeled test images providing predicted bounding boxes of its location. The predicted bounding boxes were then compared to labeled bounding boxes to calculate its Intersection-over-Union (IoU). For this model we conducted five independent trials for the dataset, each of which randomly split the dataset by patients into a training set (105 patients) and a test set (27 patients). A validation set (11 patients) is further separated from the training set to decide the training epoch. The number of training lesions (2864 +/- 26) varied depending on the random split. We counted it as a hit if a predicted bounding box had an IoU value greater than 0.5. After trying different object detection algorithms, we finally found YOLO-v5 Net [20] consistently provided the best performance. We have used YOLO-v5 Net out of the box and trained it with the default architecture. When trained and tested on the all-lesion dataset, it produced an area under the precision-recall curve (AUC-PR) of 0.69 (95% CI, 0.66-0.72) in terms of achieving an IoU>0.5 (Figure 2). By fixing the confidence value threshold to 0.3 (a value that subjectively was felt to provide a good balance between precision and recall), the recall (number predicted bounding boxes with IoU>0.5 / total number of lesions) was 0.63 (95% CI, 0.53-0.73) and precision (number predicted bounding boxes with IoU>0.5 / total number of predictions) was 0.64 (95% CI, 0.58-0.70).

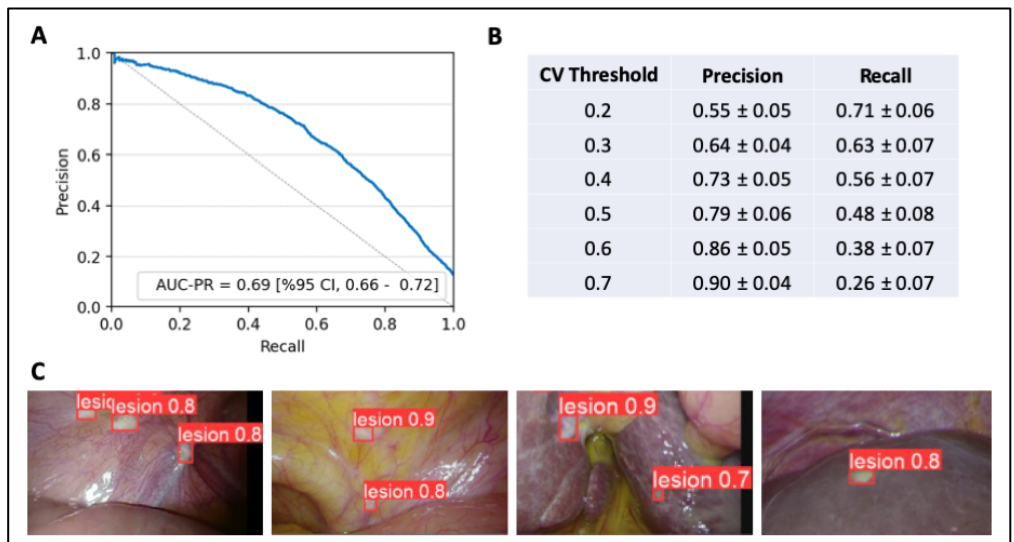

Figure 2: A) Precision-recall curve of CASL's detection system. B) CASL detection system's precision and recall dependent on the chosen confidence value threshold. C) Examples of CASL detection system's predicted lesion boundaries with confidence value.

In addition, the detection system was tested on 3650 images of 365 biopsied lesions from 132 patients (i.e. 'biopsied lesion dataset'). The AUC-PR on this dataset was 0.27 (95% CI, 0.17-0.38). The decreased performance on the biopsied lesions dataset is likely due to presence of non-biopsied lesions on the image that the system might recognize, but that do not have a manual label (i.e. providing an incorrect high number of false positive). This explains why with a confidence value threshold of 0.3, the precision was only 0.19 (95% CI, 0.14-0.24). However, the recall was 0.77 (95% CI, 0.69-0.85) suggesting that a large fraction of biopsied lesions (presumably clinically more concerning lesions) was detected correctly.

**CASL Classification System**
In the classification model, lesion patches (defined as the image within the bounding box drawn tightly around a biopsied lesions) were taken from the biopsied lesion dataset and used for training, validation, and testing. For each peritoneal surface lesion, there were 10 lesion patches available taken from different video frames providing the classification dataset of 3650 lesion patches (365 lesions, 33% metastases, 67% benign). The classification system was evaluated using 10-fold cross validation. In each training fold, there are 328 lesions in the training set and 37 lesions in the test set. The lesion patches were pre-processed by resizing them to the same size and providing histogram equalization. A wide range of DNN architectures with and without different data augmentation methods were tried. For almost all architectures, overfitting presented a problem. Finally, we used a supervised ensemble learning approach with ResNets to mitigate the overfitting issue. In each data split, we randomly further split the training set into a smaller training set and a validation set, and then trained a classifier. This was done with 30 random splits to train 30 classifiers. The top 5 classifiers with the highest validation AUC values were chosen to form the ensemble. Of note, the 30 classifiers all used different validation sets to increase the independence among the 5 chosen classifiers. To further increase the independence, each classifier was randomly choosing its architecture from a reasonable range. The training of each classifier used CutMix [21] to augment the data. In each cross-validation fold, the chosen 5 top classifiers formed an ensemble and voted for the label of a lesion patch in the test set (370 lesion patches). The system on average achieved an AUC-ROC of 0.78 (95% CI, 0.73-0.83) (Figure 3) and, when using a probability threshold of 0.5, an accuracy of 0.78 (95% CI, 0.74-0.82). The computation time was acceptable because of a relatively small training set.

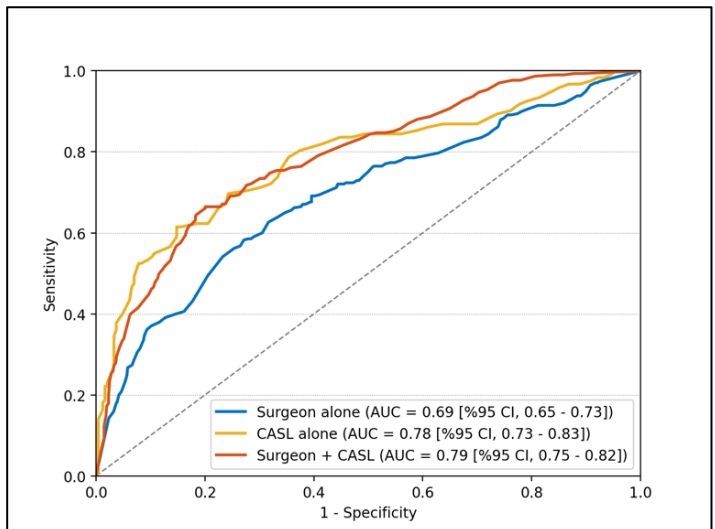

Figure 3: Receiver operating characteristic curve of predicted probability of a peritoneal surface lesion being malignant by surgeon alone (blue), CASL alone (yellow), and surgeon when supported by CASL (red).

**Human Classification Performance**
To simulate human expert performance on the same task, a national online survey evaluated oncologic surgeons' classification performance on study images from the biopsied lesion dataset without labels. Of 17,303 surveys sent out, 111 responses were received. The respondents covered a broad range of practicing oncologic surgeons. There were surgical oncologists (n=52), general surgeons (n=26), HPB surgeons (n=15), colorectal surgeons (n=11), and others (n=7). Twenty-three had been in independent practice for 0-5 years, 17 for 5-10 years, 33 for 10-20 years, and 38 for over 20 years. Forty-five were practicing in the Northeast region of the United States, 31 in the South region, 17 in the Midwest region, 12 in the West region, 0 in the Pacific region, and 6 in an unknown region. The mean number of cancer operations performed by this cohort was 12 +/- 9 per month, including 3 +/- 4 staging laparoscopies per month.
Since only biopsied lesions were used for the survey, the survey responses were compared to the underlying histopathology result. The survey provided results on 1086 peritoneal surface lesions regarding the recommendation whether the surgeon would have conducted a biopsy of the depicted lesion. The surgeons' accuracy in recommending a biopsy for metastases and omitting a biopsy for benign lesions was only 52% with a false negative rate of 21% and a false omission rate of 22% (Figure 4). Experience (number of years in practice (F(3,107)=2.532, p=0.061), number of cancer operations performed per month ($R^2$=0.014, p=0.222), number of staging laparoscopies performed per month ($R^2$<0.001, p=0.806)) did not correlate with each surgeons' accuracy. The survey provided results of 959 peritoneal surface lesions assessing the surgeon-perceived probability of a lesion being malignant. The results concluded that human expert performance provided an AUC-ROC of 0.69 (95% CI, 0.65-0.73) in correctly classifying peritoneal surface lesions (Figure 3).

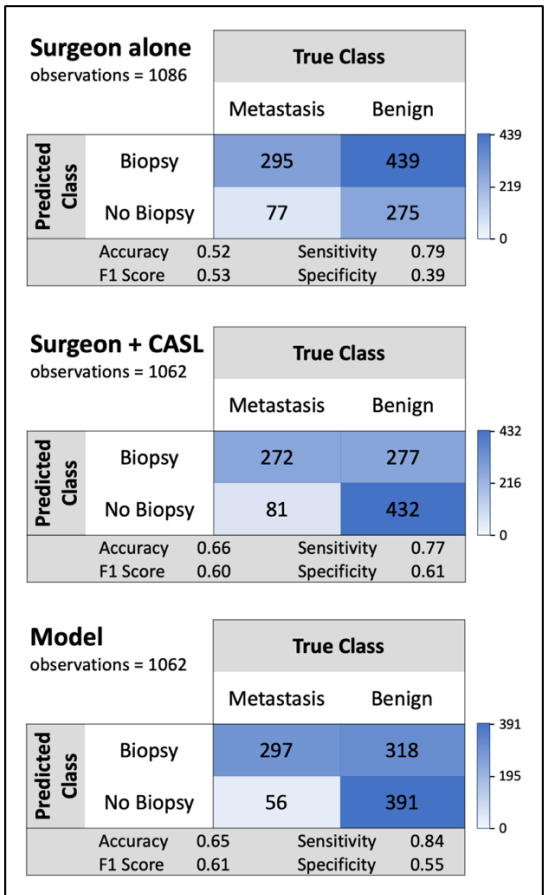

Figure 4: Survey results assessing whether surgeons would have biopsied a particular lesion in clinical practice. Confusion matrix of the performance of surgeons alone (i.e. surgeon assessing unlabeled images) (top), surgeons supported by CASL classification system (i.e. surgeon assessing labeled images) (middle), and a model combining CASL alone and surgeon supported by CASL (bottom).

**Combined Human-Computer Classification Performance**
To simulate how oncologic surgeons would perform in the classification task when supported by CASL, the same surgeons were also asked on the survey to evaluate study images from the biopsied lesion dataset with provided predictions from CASL. The results suggest that CASL improved the surgeons' performance. Specifically, according to the predicted probability of a biopsied peritoneal surface lesion being malignant, the AUC-ROC for the surgeon alone provided from the survey, CASL alone provided from the developed classification system, and surgeon when supported by CASL provided from the survey was 0.69 (95% CI, 0.65-0.73), 0.78 (95% CI, 0.73-0.83), and 0.79 (95% CI, 0.75-0.82), respectively (Figure 3). This demonstrated that the performance for classification provided by CASL alone was already better than the performance of surgeons alone (p=0.007). While the performance of surgeons alone significantly increased when supported by CASL (p=0.001), the performance of surgeons when supported by CASL was not better than the performance of CASL alone (p=0.436).

To evaluate how this improvement in AUC-ROC would correspond to changes in clinical practice, the survey's results on whether a surgeon recommended a biopsy of a particular lesion or not were further analyzed. When the oncologic surgeons evaluated images without the support from CASL compared to images that provided additional classification information from CASL (1062 images evaluated), the surgeons' accuracy in correctly recommending a biopsy increased from 52% to 66% (Figure 4, p<0.001). In detail, the sensitivity changed from 79% to 77% (p=0.537) and the specificity changed from 39% to 61% (p<0.001); indicating that the combined human-computer performance appeared to improve the identification of benign lesions, but was equivocal in identifying metastases compared to current standard practices. This led to a theoretic reduction in unnecessary biopsies by 37% (change in false negatives).

When analyzing the relationship between the predicted probability of a lesion being a metastasis provided by CASL and the mean probability provided by the surgeon for each lesion, analysis suggests a correlation but with low precision ($R^2$=0.299, p<0.001, Figure 5). The low precision likely represents the inaccuracy / error that both methods produce (Figure 6). But it might also suggest that both methods used different approaches to come to their conclusion providing an opportunity for a combined model. When assuming a hypothetical clinical practice model of performing biopsies for 1) all lesions where the surgeon when supported by CASL recommended a biopsy and 2) all lesions where CASL predicted a probability of metastasis of 0.5 or greater independent of the surgeon's recommendation, the sensitivity would have been 84% with a specificity of 55% (Figure 4). This suggests that such a model could enhance the detection of metastases by 5% (change in sensitivity) while reducing the number of unnecessary biopsies by 28% (change in false negatives) compared to current standard practice.

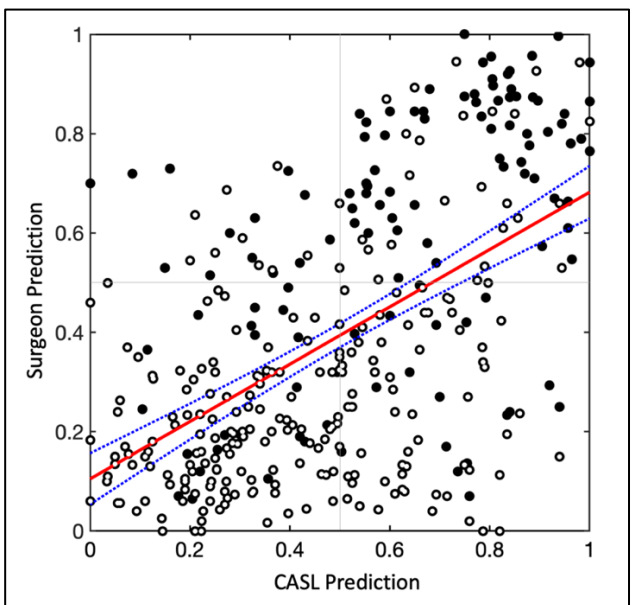

Figure 5: Mean probability of a peritoneal surface lesion representing a metastasis proposed by surgeon alone versus CASL alone. Data was available from 354 lesions (white dots – benign, black dots – metastasis, red line – linear regression trend line, blue line – 95% confidence interval).

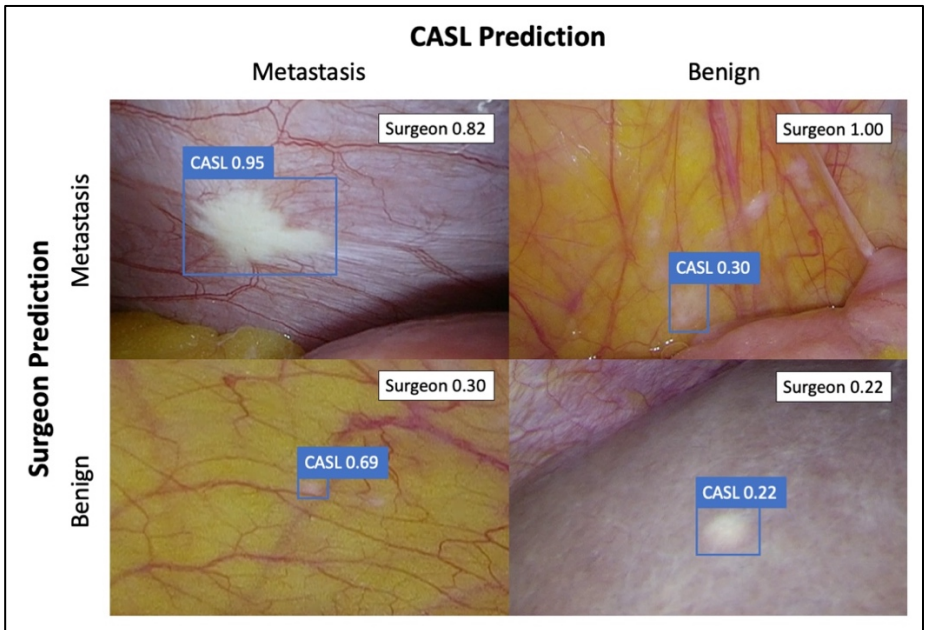

Figure 6: Examples of four biopsy-proven metastases captured during staging laparoscopy depicting different scenarios of image classification predictions.

**CASL Performance on Non-biopsied Lesions Within Patients of Presumed Localized Cancer**
To evaluate whether CASL's classification system would be capable of detecting clinically occult peritoneal surface metastases, the system was used to evaluate all visible, non-biopsied peritoneal surface lesions in patients who underwent cancer resections with curative intent in the absence of any known distant metastasis at the time of the operation. Since the ground truth (i.e. biopsy) does not exist for these lesions, CASL's classification predictions were compared to the patients' peritoneal carcinomatosis-free survival and disease-free survival as surrogates for the absence or presence of occult peritoneal metastases at the time of operation. Of 66 patients who underwent resection of their cancer with curative intent, 16 had at least one non-biopsied peritoneal surface lesion that CASL predicted the probability of metastasis to be 0.5 or greater (predicted metastasis). For the remaining 50 patients, CASL predicted the probability of metastasis of all peritoneal surface lesions to be less than 0.5 (predicted benign). In the predicted benign group, the peritoneal carcinomatosis-free survival was significantly greater compared to the predicted metastasis group (hazard ratio=3.86, p=0.044, Figure 7) with a 1-year, 2-year, and 3-year survival of 98%, 89%, and 89% in the predicted benign group compared to 81%, 73%, and 61% in the predicted metastasis group. Similar findings were made for disease-free survival (hazard ratio=2.05, p=0.098). The findings, therefore, support the possibility that CASL could identify clinically occult peritoneal surface metastases.

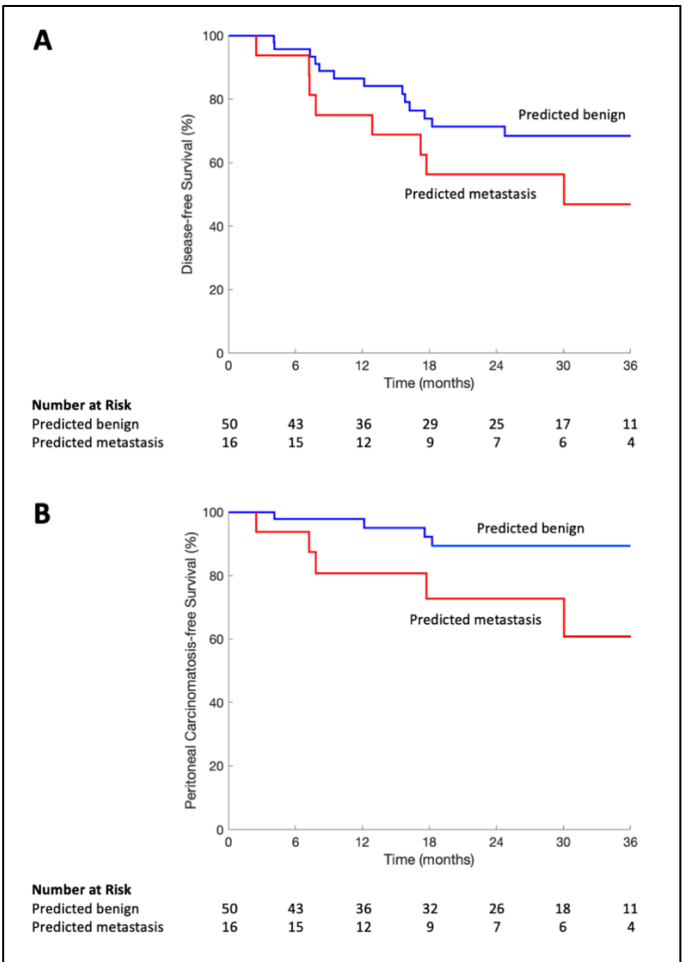

Figure 7: Kaplan-Meier estimates of disease-free survival (A) and peritoneal carcinomatosis-free survival (B) within 66 patients whose cancer was resected with curative intent (i.e. no detectable disease immediately after operation) comparing 50 patients where CASL predicted absence of metastases (predicted benign) to 16 patients where CASL predicted presence of metastases (predicted metastasis).

## Discussion

The described national survey of human experts demonstrated poor performance of oncologic surgeons in correctly identifying clearly visible peritoneal metastases (lesion classification accuracy 52%); a task that despite the absence of prior proof has historically been perceived to be reliable. In clinical practice, a false positive prediction (i.e. an unnecessary biopsy) is a much lesser problem than a false negative prediction (i.e. a missed metastasis). Therefore, the ultimate goal of any operative staging is to avoid any false negative predictions. With this in mind, the survey demonstrated that the false negative rate of human experts (i.e. the probability of not performing a biopsy of a clearly visibly metastasis) and the false omission rate (i.e. the probability that a metastasis was present when the decision was made not to biopsy) was 21% and 22%, respectively. This strongly indicates the need for substantially improvements in current operative staging. Further, the surgeons' sensitivity of 79% and specificity of 39% on this task suggests a practice to err on the side of caution and overcall metastasis, likely due to surgeons' self-recognized doubts. While there is no significant literature on this subject, the findings are in line with a prior similar, but smaller dual-institutional study by our team [16]. A hypothetical solution is to biopsy every visible peritoneal surface lesion (as some surgeons anecdotally claim they do). Our data demonstrated that each patient on average has 33 visible peritoneal surface lesions (most presumably not cancerous); making biopsy with frozen section histopathology of every lesion practically impossible and emphasizing the importance of supportive technology, such as CASL.

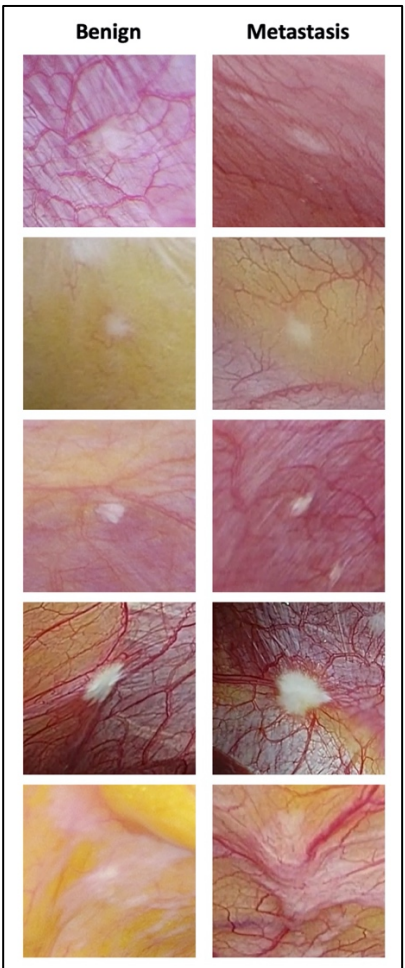

Figure 8: Examples of biopsied peritoneal surface lesions from various study patients demonstrating benign and malignant pathology with subjectively paired similar optical appearance.

The developed CASL prototype provided an encouraging performance. Its ability in detecting any peritoneal surface lesion was moderate (AUC-PR 0.69). For the detection system a comparative human survey was not conducted, since by definition the gold standard for detection of peritoneal surface lesion is the surgeon's ability to see an "abnormality". CASL's ability to classify peritoneal surface lesions was moderate (AUC-ROC 0.78); yet, still superior compared to the human experts (AUC-ROC of 0.69). Like in many other scenarios where computer vision is used, we also observed suggestions of presumed different approaches between humans and computers in providing a prediction. These differences come with a theoretic potential for complementary effects when human and computer work together, which can lead to even better results compared to each performance on their own given. However, when surgeons use the adjunct of CASL in a simulated clinical environment, the AUC-ROC for classification improved significantly compared to surgeons alone, but was not better compared to CASL alone (surgeon alone 0.69, CASL alone 0.78, combined 0.79). Additional training of surgeons and development of trust in the CASL system are potential future pathways to further increase the combined performance. If CASL would be applied in clinical practice under set rules, at least in the simulated clinical environment, it has the potential to enhance the detection of metastases by 5% while reducing the number of unnecessary biopsies by 28% compared to current standard practice. In addition, the ability for CASL to detect clearly visible, yet clinically occult, metastases was further demonstrated by the observed difference in cancer-free survival among patients who underwent resections with curative intent where CASL predicted at least one hidden metastasis versus similar patients where CASL predicted no metastases. Yet, despite this promise, significant further development of this prototype will be needed before considering a pivotal trial. The overall challenge in this task is the heterogeneity in optical appearance of benign and malignant peritoneal lesions with significant overlap in appearance [16]. Something a machine learning model frequently is more capable to overcome than human visual perception. But it does require additional data for training of the system.

This study supports the feasibility of using deep learning algorithms for operative staging; an approach that previously has never been applied in this domain. The challenge in the task is the high prevalence of peritoneal surface lesions in patients (33 per patient in this study, most of them presumed to be benign) and the striking similarity between benign lesions and metastases [16] (Figure 8). Benign lesions can be a result of many pathologies, including fibrosis, mesothelial hyperplasia, fat necrosis, etc. The commonality of benign lesions and metastasis is the abundance of collagen [22], making their gross optical appearance similar and therefore difficult to differentiate. Nevertheless, deep learning has the potential to provide classification combining subtle changes in color, texture, and morphology as demonstrated in this study. This technology has the added potential of being used for 1) many other cancers in the abdomen and chest; 2) detection of early, small, otherwise occult peritoneal surface metastases; and 3) identification of metastases in a low resource environment where either frozen section histopathology or specialty surgical expertise might not be available. The envisioned end product of this prototype is an automated surgical guidance system that uses standard laparoscopy imaging and is capable to identify and classify peritoneal surface lesions at video rate along with an output label displayed to the surgeon through an augmented reality display in real time. Such a system would direct surgeons to any location requiring biopsy, or if future performance of the system is outstanding, eliminate the need for biopsy completely. A theoretic alternative approach to our label-free imaging technique is bio-labeled operative imaging, such as fluorescence-guided cancer surgery, which has been tested in pivotal trials for anti-folate receptor and anti-carcinoembryonic antigen-labeled agents identifying additional peritoneal metastases in 33% and 16% of patients, respectively [23-25]. Yet, the need for such imaging agents has inherent limitations of washout, limited use to certain cancer types, timing of application at point of care, and for some toxicity and photosensitivity [23]; suggesting that a label-free approach is advantageous. Future implementation of CASL only requires the addition of a graphics processing unit or application-specific integrated circuits, which are commercially available and can be incorporated into existing laparoscopy processors.

Given the current prototype phase of the system, limitations of the study exist. The current CASL system was developed on still images only, detection and classification algorithms were at this stage not combined yet, and human performance was assessed in a simulated clinical environment (i.e. survey). Validation of the results in true clinical practice along with CASL running at video rate still has to occur. Further, performance was assessed on a lesion level. How this will influence performance on a patient level where there might be more than one peritoneal surface lesion visible and, therefore, potentially more than one metastasis that could be correctly identified remains unknown. Results on a patient level could not reliably be determined in this study given the selective biopsy of representative lesions (i.e. a surgeon in theory could have misclassified a metastasis on the survey from a biopsied lesion, but in real practice would still have made the correct diagnosis by biopsy of a different metastasis if present). The classification sample only included peritoneal surface lesions that were biopsied in accordance to a single surgeon's practice, but does not include lesions that would have been biopsied by other surgeons and does not exclude lesions that would not have been biopsied by any other surgeon. This introduces the potential for selection bias and overfitting of the model requiring future validation in a multi-institutional study. The survey had a low response rate, but all responding surgeons were by definition qualified examiners minimizing the potential for non-response bias.

In summary, failure to notice grossly visibly peritoneal surface metastases is likely a major reason for poor oncologic outcome in a significant number of patients who undergo operative treatment for various gastrointestinal cancers. A concept that is expected to also apply to many non-gastrointestinal cancers. Since benign peritoneal lesions are abundant with frequently similar appearance compared to peritoneal surface metastases, the demonstrated high rate of misclassification of metastases by surgeons was not surprising. The resulting poor performance by surgeons was counteracted by CASL as demonstrated in a simulated environment where CASL was able to identify metastases that were missed by human examiners. Despite the obvious need for further development and refinement of the presented prototype system and future testing in a multi-institutional clinical validation study with a fully combined system that functions at video rate, this study demonstrated the technical feasibility of the system and the tremendous potential to improve clinical care decisions for a large number of surgical cancer patients. Future development of the system, beyond the addition of more samples, might include adding other clinical features to the system such as the number of lesions seen in a field of view or whether ascites is present or not. Also, upcoming development will have to take into consideration the clinical implications of identifying otherwise occult metastases. Specifically, as any cancer detection method becomes increasingly sensitive, it can become difficult to distinguish metastases that are practically inconsequential from those that are life-threatening [26]. While more information is expected to lead to better treatment allocation and thereby better outcomes, progress on better understanding individualized risk and implementation of better treatment pathways for oligometastatic disease is going to be needed. CASL is expected to help in this endeavor.

# Methods

### Data Source
Conventional staging laparoscopy videos were obtained from an established image library containing a series of operative videos performed by a single surgeon (T.S.) under the standard of care between January 2014 and July 2021. Inclusion criteria encompassed: 1) adult patients (≥18 years of age), 2) underwent a staging laparoscopy as part of routine treatment for histologically confirmed adenocarcinoma involving the gastrointestinal tract, 3) patient signed an operative consent form that includes consent for video recording, 4) functional video recording was available depicting the entire length of the staging laparoscopy, 5) at least one biopsy of a peritoneal surface lesion was performed at the time of operation. Individuals involuntarily detained in a penal institution or who have impaired capacity to make informed medical decisions were excluded.

Relevant clinical data was obtained from the patient's electronic medical record. All available histopathology slides of biopsied peritoneal surface lesions were re-reviewed by a single gastrointestinal pathologist (A.G.N.). Discrepancies with the original pathology result were resolved by group consensus. Operative videos were reviewed to assure that the performed biopsy was considered good quality with the pathology specimen well representing the peritoneal lesion. Any individual lesion with indeterminate pathology result was excluded from the analysis. The study was approved by the Lahey Clinic IRB.

### Preprocessing and Annotation of Data
For development of the detection system, still images best representing each visible peritoneal surface lesion (biopsied and non-biopsied) were captured from each staging laparoscopy video by a trained annotation specialist (J.C.). More than one lesion per image was allowed, but duplicates of lesions were eliminated. Each image underwent labeling by manually placing a bounding box tightly around each lesion. The unlabeled and labeled images of each lesion represented the data for supervised deep learning. The data, therefore, provided a single representative full-view laparoscopy image of each visible peritoneal surface lesion with a label of its location within the image.

For development of the classification system, 10 still images best representing each biopsied peritoneal surface lesion from different angles and distances were obtained from each video. A sample size of 10 images per lesion was chosen in order to get as great of a number of images as possible without duplicating the point of view. Each image underwent cropping to create an image patch that tightly included each biopsied lesion. The image patches of each biopsied lesion and its histopathology results ('benign' vs. 'metastasis') represented the data for supervised deep learning. The data, therefore, provided 10 image patches of each biopsied peritoneal surface lesion with a label of the underlying pathology.

All images had the overarching goal of obtaining a clear / non-obstructed view, with good illumination, good focus, and good resolution. A peritoneal surface lesion was defined as a limited area of abnormal appearing tissue on the parietal or visceral peritoneum clinically concerning for the possibility of metastasis. Each image was stored in PNG format (24-bit RGB, lossless compression). All images underwent quality review by a single surgeon (T.S.). A total of 365 biopsied and 3922 non-biopsied peritoneal surface lesions were identified from 132 staging laparoscopy videos providing 1092 images depicting 4287 lesions for development of the detection algorithm and 3650 image patches for development of the classification algorithm.

### Development of the Deep Neural Network
*Detection System*
The detection system was designed to autonomously place a tight bounding box around any peritoneal surface lesion area within a laparoscopic still image through training of a DNN. Various DNNs with different architectures were trained using supervised learning. The detection images were divided into a training and a test set using five random splits on a patient-level to avoid data leakage. In each random split, a small validation set is further separated from the training set to monitor the training process and decide the number of training iterations. On each test image, the system predicted one or multiple bounding boxes on unlabeled images with the goal to predict the location of a lesion. Each predicted bounding box was associated with a confidence score in the range of 0.0 to 1.0 expressing the system's confidence that there is a lesion in the predicted bounding box. Every predicted bounding box was compared with its nearest manually labeled bounding box on the labeled images to compute the IoU (i.e. the proportion of overlap between the predicted and the manually marked bounding box). An IoU of 0.5 or greater was considered a hit, allowing the calculation of a confusion matrix at various confidence score thresholds (i.e. IoU >= 0.5 and confidence score above threshold = true positive; IoU < 0.5 and confidence score above threshold = false positive; IoU < 0.5 and confidence score below threshold = true negative; IoU >= 0.5 and confidence score below threshold = false negative). The detection system's performance was measured by AUC-PR under varying thresholds of the system's confidence score.

*Classification System*
Classification of peritoneal surface lesions into benign lesions or metastases occurred by training of different supervised DNNs using image patches as data source. The evaluated systems applied different data augmentation methods and different architectures to provide a prediction in the range of 0.0 to 1.0 about the probability of a lesion representing a metastasis. The image patches were divided into a training / validation set and a testing set using a split on lesion level to avoid data leakage. The system's predicted probability of the presence of a metastasis was compared to the underlying histopathology result on biopsy. Performance on the testing set was measured by the area under the receiver operating characteristic curve (AUC-ROC) under varying thresholds of the system's computed probability.

**Simulated Clinical Evaluation of the Deep Neural Network**
*Performance Comparison of Classification System*
A national online survey was conducted asking oncologic surgeons to assess study images without labels and study images with provided predictions from CASL. The performance of the surgeon in classifying peritoneal surface lesions was recorded using Qualtrics XM software (Qualtrics, Provo, UT). To develop the survey, from each of the 365 biopsied peritoneal surface lesions, one representative full-view laparoscopy image was chosen in order to create an unlabeled and a labeled sample. The unlabeled sample included 365 images with the addition of a visible arrow marking the location of the lesion to be evaluated. The labeled sample included the same 365 images from the unlabeled sample yet without the arrow but instead annotated by a bounding box around the lesion and a label citing the probability of metastasis computed by the CASL classification algorithm. The surveyed oncologic surgeons were provided a sequence of 10 random unlabeled images followed by 10 random labeled images. All 20 images represented different peritoneal surface lesions. To maintain an equal number of evaluations per lesion, block randomization was used. The surgeons were aware that the images originated from staging laparoscopies performed on cancer patients, but they were blinded towards the pathology result, the fact that all lesions had been biopsied, and any other clinical information. The surgeons were aware that the provided probability estimates on the labeled images originated from an artificial intelligence system, but were blinded towards any known performance measures of the system. The surgeons were asked to grade the probability of each peritoneal surface lesion representing a metastasis on a continuous sliding scale between 0 and 100. In addition, they were asked to determine whether they would have biopsied such a lesion in their clinical practice (i.e. 'yes'/'no'). The results were used to compare the performance of surgeons alone (i.e. surgeon performance on unlabeled images) versus CASL alone versus surgeons when supported by CASL (i.e. surgeon performance on labeled images) in this simulated clinical environment.

Email contact information was obtained from the American Medical Association Physician Masterfile, a comprehensive record of all US physicians independent of American Medical Association membership. All surgeons with the listed specialty of abdominal surgery, advanced surgical oncology, general surgery, gynecological oncology, and surgical oncology with an office- or hospital-based practice within the United States were chosen. The sample included the contacts of 17,303 surgeons, who received an invitation email and a single reminder email. The survey was conducted between April 12, 2022 and June 19, 2022.

The CASL detection algorithm was not used in the survey, since its purpose was to facilitate the CASL classification algorithm. Further, the ground truth for the detection algorithm is provided by a human examiner and, therefore, surgeons for the most part are expected to perform as well or better in localizing peritoneal surface lesions than the detection algorithm.

*Identification of Clinically Missed Peritoneal Metastases*
To test the potential for CASL to identify peritoneal surface metastases that were missed during routine clinical care, the performance of CASL's classification system on images from patients that were clinically not deemed to have distant metastases at the time of their operation and underwent a resection of the underlying cancer with curative intent (i.e. margin negative resection) was assessed. Specifically, from 66 patients without known metastases at the time of operation all available labeled images from the dataset used for development of the detection system were obtained. Any biopsied lesions previously used to train the classification system were excluded to avoid data leakage. A total of 568 images depicting 1518 non-biopsied lesions from 66 videos were utilized. Image patches of each labeled peritoneal surface lesion were created (i.e. one patch per lesion) and subsequently assessed by CASL's classification system. The probability of a metastasis was computed by CASL's classification system for all 1518 provide image patches. Each patient was assigned its maximal probability score derived from the patient's lesions. Information on the date and location of radiographic cancer recurrence and survival status was obtained from the patients' electronic medical record. CASL's maximal predicted probability of a metastasis being present on a patient-level was then compared to the observed cancer-free survival.

**Computational Environment and Statistical Analysis**

Training, validation, and testing of the systems were carried out on a 32-core workstation equipped with two V-100 GPUs. Classification AUC-ROC was calculated amongst surgeon, CASL, and combined performance using probability thresholds. For the AUC analysis, surgeon performance was generated from all survey probability evaluations (each lesion generated multiple evaluations) and CASL performance was computed from each lesion (each lesion produces one aggregate result). Statistical analyses to compare the AUC amongst surgeon, CASL, and combined performance used a paired test through bootstrapping method. For comparison of the confusion matrixes amongst surgeon, CASL, and combined performance, Chi-square test was used. For comparison of the surgeons' experience and surgeons' accuracy, linear regression was used. For patient survival analysis, Kaplan-Meier estimates with a Cox proportional hazards model for group comparison was used. All hypotheses tests were 2-sided and a p-value less than 0.05 was considered significant.